# Defining, measuring, and modeling passenger's in-vehicle experience and acceptance of automated vehicles

Neeraja Bhide* [a], Nanami Hashimoto* [b], Kazimierz Dokurno* [c], Chris Van der Hoorn* [a], Sascha Hoogendoorn-Lanser [d], Sina Nordhoff [d]

[a] Delft Center for Systems and Control, Delft University of Technology, The Netherlands
[b] Department of Cognitive Robotics, Delft University of Technology, The Netherlands
[c] Faculty of Electrical Engineering, Mathematics and Computer Science, University of Twente, The Netherlands
[d] Department Transport & Planning, Delft University of Technology, The Netherlands

* Corresponding author

**ABSTRACT**

Automated vehicle acceptance (AVA) has been measured mostly subjectively by questionnaires and interviews, with a main focus on drivers inside automated vehicles (AVs). To ensure that AVs are widely accepted by the public, ensuring the acceptance by both drivers and passengers is key. The in-vehicle experience of passengers will determine the extent to which AVs will be accepted by passengers. A comprehensive understanding of potential assessment methods to measure the passenger experience in AVs is needed to improve the in-vehicle experience of passengers and thereby the acceptance. The present work provides an overview of assessment methods that were used to measure a driver's behavior, and cognitive and emotional states during (automated) driving. The results of the review have shown that these assessment methods can be classified by type of data-collection method (e.g., questionnaires, interviews, direct input devices, sensors), object of their measurement (i.e., perception, behavior, state), time of measurement, and degree of objectivity of the data collected. A conceptual model synthesizes the results of the literature review, formulating relationships between the factors constituting the in-vehicle experience and AVA acceptance. It is theorized that the in-vehicle experience influences the intention to use, with intention to use serving as predictor of actual use. The model also formulates relationships between actual use and well-being. A combined approach of using both subjective and objective assessment methods is needed to provide more accurate estimates for AVA, and advance the uptake and use of AVs.

# 1. Introduction

With the latest technological advances in Artificial Intelligence and microprocessor technology, Advanced Driver Assistance Systems (ADASs) are being developed into higher levels of automation with the promise of maximizing the experience for its users, contributing to the realization of societal and economic benefits, such as increased road safety, energy efficiency, and productivity (Beggiato, Hartwich, & Krems, 2018; Kyriakidis, Happee, & De Winter, 2015; Pascale et al., 2021). The scientific community has witnessed an increasing number of research studies that have been dedicated to understanding automated vehicle acceptance (AVA) (Hulse, Xie, & Galea, 2018; Kyriakidis et al., 2015).

Most studies investigating AVA focused on drivers of these automated vehicles (AVs). In SAE Level 1–3 AVs (SAE International, 2018), drivers are required to supervise the driving task and take over control to respond to critical objects and events in the environment. Studies have started to investigate the AVA by passengers (Pascale et al., 2021). Passengers can ride together with drivers, or can be transported from A to B in driverless automated passenger vehicles. Drivers of AVs reported to disengage the automation due to the discomfort of their passengers (Nordhoff & De Winter, 2023). The potential of AVs remains unexploited if the technology is not fully accepted by both drivers and passengers. An alternative negative implication is that the single vehicle miles travelled may increase if drivers decide to drive alone without passengers to experience and test the system. Ensuring the acceptance of AVs by both drivers and passengers is essential to contribute to a large-scale uptake and use of AVs.

How passengers in AVs experience the ride is vital for establishing the acceptance of AVs by passengers. Nordhoff et al. (2023) posited that the AVA by passengers is influenced by safety and trust, efficiency, (motion-)comfort, including motion sickness, pleasure, and social influence. Safety and trust, comfort, and pleasure constitute the factors impacting the in-vehicle experience, which corresponds with Dillen et al. (2020). In comparison to drivers actively controlling the driving task, the perceived risk may be higher for passengers as passengers may not understand the intentions and actions of AVs, and cannot intervene in the driving task (Pascale et al., 2021). Therefore, a person's locus of control will be pivotal for a passenger's perception of the in-vehicle experience and acceptance (see Dillen et al., 2020). In line with technology acceptance models (Venkatesh, Thong, & Xu, 2012), individual differences, such as age and gender, influence the strength of the relationship between the in-vehicle experience and acceptance. Further, we posit that the intention to use influences actual use. Actual use of AVs, in turn, impacts a person's well-being. The impact of travel mode choice on a person's well-being has been established (Singleton, 2019). We postulate that AV behavior and the environment serve as external variables impacting the in-vehicle experience (see Davis, 1985).

In previous studies, AVA was mostly assessed subjectively, e.g., by questionnaires and interviews. Common biases exist with self-reported data collected from questionnaire and interview research, some of which were also evidenced in studies investigating AVA:

- Social desirability bias: Respondents providing responses in a socially desirable rather than in an accurate and truthful way (Holtgraves, 2004). Possible reasons could be the novelty factor, the influence of the media in marketing AVs (Nordhoff et al., 2020; Palatinus et al., 2022), or the fear of consequences of disclosing personal sensitive information (Habibifar & Salmanzadeh, 2022);
- Cognitive dissonance bias: Inconsistency between a person's cognitive state. When a person's previous actions and their beliefs do not align, a negative emotional reaction can occur. In order to avoid this cognitive dissonance, people are known to adjust their viewpoints or behaviour to better align (Wu, Lin, & Liu, 2020);
- Hypothetical bias: Respondents rating AVs they have not physically experienced (Palatinus et al., 2022);
- Misunderstanding of the questionnaire instructions and items: Respondents misunderstanding the meaning of the questionnaire instructions (e.g., description of AVs), and items. For example, the questionnaire item 'Using an automated car would be useful in my daily life' (Nordhoff et al., 2020) is formulated in very generic terms, which may lead respondents to attach a different meaning to this item;
- Inaccuracy: Respondents not accurately rating their agreement with the questions, e.g., an example can be found with regards to respondents' cognitive and emotional states (Palatinus et al., 2022);

- Response style (or response bias): Respondents' tendency to respond in a certain way regardless of the question. Extreme response bias refers to a tendency to respond with an extremely low or high rating score. Response behavior providing high rating scores can be referred to as acquiescence bias or 'yea-saying behavior' (Jackson & Messick, 1958; Van Vaerenbergh & Thomas, 2013), which was also observed in prior AVA studies (Nordhoff et al., 2018). Moderate bias, in contrast, refers to a tendency to give medium ratings for all questions. Acquiescence bias and disacquiescence bias refer to a tendency to respond in an overly positive or negative manner (Jackson & Messick, 1958; Van Vaerenbergh & Thomas, 2013).

The development of biosensing technologies has made it possible to investigate AVA more objectively, collecting both behavioral and physiological data (e.g., HR, eye gaze movements, galvanic skin response (GSR), facial expressions). Behavioral and physiological measures are commonly used to assess driver's cognitive state, e.g., vigilance, workload, attention, fatigue and drowsiness (Biondi et al., 2018; Lohani, Payne, & Strayer, 2019). Findings from those studies indicate that the accuracy of detecting driver's cognitive state can be increased by utilizing physiological and behavioral measures (Lenneman & Backs, 2009; Yang, Hosseiny, Susindar, & Ferris, 2016). The use of more objective measures for the assessment of AVA is still in its infancy (Palatinus et al., 2022). Physiological and behavioral measures of AVA can reduce the subjectivity of self-reported data collected in questionnaire and interview research. The results obtained from these measures can be compared to the self-reported data (Palatinus et al., 2022), helping researchers and practitioners to improve their understanding of AVA. However, similar to subjective assessment methods, objective assessment methods have several caveats, such as:

- Limitations in sensor technology, contributing to inaccurate and inconsistent measurements. For example, the wristwatches used in the study of Pascale et al. (2021) did not measure respondents' heart rate consistently, which contributed to excluding the data from the study. Beggiato et al. (2018) referred to the differences in the sensitivity of the devices to measure skin conductance (e.g., looseness of the band, sensitivity of hand movements), which might account for the non-significant relationship between discomfort and skin conductance.

- High sensitivity towards individual differences (Lal & Craig, 2001). For example, in the study of Irmak, Pool, and Happee (2021) the nervousness of respondents might have affected the measurement of GSR. Hu et al. (2016) mentioned that GSR is influenced by muscle contraction, which makes the measurement of GSR less suitable if the task requires movement. Seet et al. (2020) pointed out that trust could be accurately measured as their sample had circumscribed demographics in terms of age and neurotypicality. In the study of Beggiato, Hartwich, and Krems (2019) the unexpected decrease in HR as measure for discomfort was explained as an effect of "preparation for action", which represents an anticipatory deceleration of HR prior to planned actions.
- High sensitivity towards experimental conditions: Experimental conditions (Hu, Akash, Jain, & Reid, 2016; Irmak, Pool, et al., 2021; Palinko, Kun, Shyrokov, & Heeman, 2010) can confound the physiological measurements. For example, in the studies of Hu et al. (2016) and Irmak, Pool, et al. (2021), the temperature and humidity might have affected the measurement of GSR. Palinko et al. (2010) have shown that the eye pupil size was strongly dependant on ambient light.
- Low effectiveness: High methodological preparedness and time for the collection of physiological and behavioral data is needed to achieve a sufficiently large sample size compared to survey research (Lukovics et al., 2023).

This paper provides an overview of assessment methods used to measure driver's cognitive and emotional state during manual and automated driving. This review has the following three main contributions. First, it provides insights into more objective measures of AVA, and their relationship with the subjective self-reports, as well as their effectiveness in measuring, explaining, and predicting AVA by passengers in comparison to the subjective self-reported measures.

## 2. Assessment methods

Assessment methods are data collection methods that can be broadly classified by the degree of:
- Type of assessment method: How is the data collected? The most common assessment methods were questionnaires, and interviews, direct input devices, photo and video observations, wearable devices and sensors.

- Object of the measurement: What do these methods measure? The object of the measurement were perceptions, behavior, and cognitive and affective psycho-physiological states. Note that perceptions can also include expectations, attitudes, or opinions. We refer to perceptions throughout the manuscript.
- Degree of objectivity of the data collected: How 'objective' is the data?
- Time of data collection: When is the data collected? Data can be collected before, after, or during the ride with an AV.

In the following section, we will focus on the types of assessment methods used to measure respondent's perceptions, behavior, and state.

## 2.1. Questionnaires

Questionnaires are subjective assessment methods, i.e., the data collected by these instruments relies on the subjective self-reported data collected from respondents. Questionnaires collect data at the moment determined by the experimenter, typically before and / or after the experience, or without having the person subjected to any specific experience at all (Nordhoff et al., 2018; Pascale et al., 2021; Bansal, Kockelman, & Singh, 2016). Questionnaires are still the most common method for assessing AVA (Palatinus, Volosin, Csabi, et al., 2022). Questionnaires are considered a cost-effective data collection method that can be distributed to a large number of respondents.

Questionnaires were typically used to measure an respondent's perception towards and acceptance of AVs applying constructs from common technology acceptance models, but can also be used to measure behavior and physiology, such as motion sickness. Questionnaires are also common to measure respondent's cognitive and affective evaluations of AVs, such as trust, perceived safety, comfort, and locus of control. For example, the 'trust in automation' scale by Jian, Bisantz, & Drury (2000), which measured trust by 12 items (5 of these items measured distrust, and 7 items measured trust on a 7-point Likert rating scale), was applied in the studies of e.g., Bellem, Thiel, Schrauf, & Krems (2018), and Dettmann et al. (2021) to assess the passenger experience in an AV. He, Stapel, Wang, & Happee (2022) and Pascale et al. (2021) developed a questionnaire to measure perceived safety of respondents' experiences with specific aspects of AV behavior. The scale consisted of items pertaining to stressfulness and road monitoring, and the perceived risk at different sections of the road, measured on a sliding scale ranging from 0 to 100. To measure ride comfort, the standard SAEJ1060 (SAE International, 2014) was developed, which provides a subjective measure of comfort on a scale from 1 to 10 (1-4 = unacceptable, 5 = neutral, 6-10 = acceptable). Informed by the 'theory of safety margins' by Summala (2007), Siebert, Oehl, Höger, & Pfister (2013) developed the 14-item - "Disco-scale" to assess discomfort in automated driving with items pertaining to situational safety margins and system components measured on a 5-point Likert scale. For the measurement of motion sickness, misery rating scales, such as the misery rating scale (MISC) (Bos, de Vries, van Emmerik, & Groen, 2010), and motion sickness susceptibility questionnaire (MSSQ) were often used (Golding, 2006). The traffic locus of control (T-LOC) scale (Özkan & Lajunen, 2005) is a well-known scale for assessing driver's locus of control. An adapted version of this questionnaire was applied by Syahrivar et al. (2021), which consisted of 34 items, with 8 items measuring driver's locus of control, and 9 items measuring desire for control, measured on a 5-point Likert scale. Another scale used to measure locus of control is the driving internality and externality scale by Montag & Comrey (1987) adapted by Choi and Ji (2015) to assess respondents driving-related internal and external locus of control personality traits. The results showed that locus of control has strong influence on the people's attitude towards AVs.

## 2.2. Direct input devices

Direct input devices, such as push buttons or joysticks, are examples to measure respondents' experiences continuously during the experiment (Beggiato, Hartwich, & Krems, 2019; Biondi et al., 2018; Nordhoff, Stapel, et al., 2020; Rubagotti, Tusseyeva, Baltabayeva, Summers, & Sandygulova, 2022). Thus, unlike questionnaires, direct input devices provide "real-time" measurements. They were typically used to measure trust and comfort (Rubagotti et al., 2022), and affective states. For example, the Affective-State Reporting Device (ASRD) by Zoghbi, Croft, Kulić, and Van der Loos (2009), which represents an in-house developed modified joystick measured respondent's affective states. Another handheld-comfort level indicator device by Koay, Walters, and Dautenhahn (2005), and Koay et al. (2006) allowed respondents to indicate their comfort level during the experiment. Other studies (Beggiato et al., 2018; Beggiato et al., 2020; Dettmann et al., 2021) measured passenger's comfort during automated driving continuously using a handset control. The literature addressing the use of direct input devices to quantify passenger experience inside AVs is limited and mainly focuses on measuring respondent's trust and comfort levels.

### 2.3. Photo & video observations, wearable devices & sensors

Data from videos and photos were collected and wearable devices and sensors (e.g., body trackers, pressure mats, thumb presses) used to measure respondent's behavior such as body movements and facial expressions (Rubagotti, Tusseyeva, Baltabayeva, Summers, & Sandygulova, 2022) (Beggiato et al., 2018; Irmak, de Winkel, Pool, Bülthoff, & Happee, 2021). Movements in eyes, hands, and feet were associated with cognitive and affect states (Palatinus, Volosin, Csabi et al., 2022). In Beggiato et al. (2018), a marker-based motion tracking system as well as a seat pressure mat were applied to measure body movements (e.g., shoulder position). The posture dynamics were strongly related to specific situations of comfort. Beggiato, Rauh, and Krems (2020) provided scientific evidence for using facial expressions as predictor for comfort during automated driving. To measure respondent's psycho-physiological states (Palatinus et al., 2022), measurements of the heart rate activity, eye gaze behavior, galvanic skin response, brain activity, and respiratory rate were commonly performed using wearable devices and sensors.

#### 2.3.1. Heart rate activity

Heart rate activity is assessed by heart rate (HR) measured in beat per minute and heart rate variability (HRV) measured in milliseconds. Heart rate activity can be measured by either electrocardiography (ECG), sphygmomanometer (i.e., measure for blood pressure), phonocardiography (i.e., measure for sound of pulse), or photoplethysmography (PPG) (i.e., measure for blood volume changes) (Arakawa, 2021). Wearable wrist watches using PPG sensors are becoming more popular because of their easily available and applicable features (Arakawa, 2021; Beggiato et al., 2018). Heart rate activity has been widely known as an indicator for stress (Healey & Picard, 2005; Kim, Cheon, Bai, Lee, & Koo, 2018; Parsi, O'Callaghan, & Lemley, 2023), and cognitive workload (Mehler, Reimer, & Coughlin, 2012; Mehler, Reimer, Coughlin, & Dusek, 2009; Yu, Bao, Zhang, Sullivan, & Flannagan, 2021). An increase in HR and a decrease in HRV were typically associated with higher mental workload and stress (Persson, Jonasson, Fredriksson, Wiklund, & Ahlström, 2020). To measure HRV, frequency domain analysis was commonly used along with the time domain analysis because changes in frequency bands were associated with certain physiological phenomena. For example, high frequency (HF) and low frequency (LF) ratio was the most common indicator for a balance of the sympathetic and parasympathetic activity. In the context of automated driving, HF and LF power were associated with self-reported motion sickness (Holmes & Griffin, 2001). Karjanto et al. (2022) showed that passengers' HR increased and HRV decreased during the ride in a mimicked automated vehicle compared to the resting condition. The difference in HR and HRV between the defensive (lower acceleration range) and assertive (higher acceleration range) driving conditions was not significant, while the self-reported discomfort increased in the assertive driving condition. Beggiato et al. (2019) showed that HR and HRV as measure for discomfort decreased during automated driving, which was an unexpected finding given that higher stress was typically associated with a higher HR. Stephenson et al. (2020) did not show any statically significant difference in HR and HRV as a measure for trust between unexpected and expected stops during automated driving. A relationship between changes in HR and perceived safety was found by Stapel, Happee, Christoph, van Nes, and Martens (2022) and He et al. (2022) in a Level-2 automation on-road study. However, no correlation between risk level and heart rate activity was evidenced by He et al. (2022).

### 2.3.2. Eye gaze behavior

Eye-tracking research has a long research tradition and has been mainly used to investigate visual attention (Duchowski, 2002), cognitive workload (Palinko & Kun, 2012; Palinko et al., 2010), and emotional arousal (Skaramagkas et al., 2021). Eye tracking data is commonly collected with tracking glasses. Chapman and Underwood (1998) established a link between road traffic accidents and deficiencies in visual attention of the driver. They found that respondents who were asked to watch video recordings of dangerous traffic events generally exhibited longer fixation durations, a reduced area of visual search and a reduced variance of fixation locations compared to people that watched safe traffic situations. Palinko et al. (2010) have shown that pupil dilation is linked to increased mental effort, information processing, fatigue and stress. Cognitive workload can also lead to a reduced mean fixation duration of glances and an increase in the sampling rate. Stephenson et al. (2020) showed an increased number of fixation and longer fixation time during unexpected stops of an AV. This result might be explained by an increase in anxiety, contributing to a narrowing of visual attention and focusing on the hazard. Beggiato et al. (2018) demonstrated that an increase in pupil diameter can be considered an indicator for passenger discomfort inside an AV. The pupil diameter can also reflect the level of perceived risk by a person if the noticed events are sufficiently risky (He et al., 2022).

### 2.3.3. Galvanic skin response

Galvanic skin response (GSR), also called electrodermal activity (EDA), is reflective of respondent's mental state, such as attention, stress (Healey & Picard, 2005), anxiety, and cognitive load (Mehler et al., 2012; Mehler et al., 2009). The GSR has two components: The tonic GSR and phasic GSR. Tonic GSR is responsible for slow changes in the GSR singal. The main part of this is the skin conductance level (SCL). Phasic GST is responsible for more rapid changes known as skin conductance responses (SCRs). An increased mental arousal stimulates the swear gland activity, increasing the skin conductance level (SCL). Irmak, de Winkel, et al. (2021) revealed a clear correlation between SCL and motion sickness assessed by the MISC scale. Increase in both tonic and phasic GSR were observed. Beggiato et al. (2018) reported no significant relationship between SCL and discomfort during automated driving in a simulator. In the study by Morris, Erno, and Pilcher (2017), respondents showed significantly higher SCL when being driven in the 'risky' automated driving mode than in the 'safe' automated driving mode.

### 2.3.4. Brain activity

Electroencephalography (EEG) has been commonly used to measure electrical activity in the brain from the scalp using multiple electrodes. The EEG measurements are classified in terms of their frequency bands, ranging from a low frequency delta band (below 4 Hz) to a high frequency gamma band (higher than 20 Hz). The frequency bands include the delta band (lowest frequency band of below 4 Hz), theta band (frequency between 4 and 8 Hz), alpha band (frequency between 8 and 12 Hz), beta band (frequency between 13 and 30 Hz), and gamma band (frequency higher than 30 Hz) (Palatinus et al., 2022). To measure cognitive and affective states, usually the ratio between high (beta) and low (alpha) frequency bands between the left and right hemispheres are compared. The substraction of left frontal alpha power from right frontal alpha power results in the condition of frontal alpha symmetry. The result of this substraction is referred to as relative left or relative right frontal activity. A more positive result means that relative left activation is present, and when the result is more negative, relative right activation is present (Palatinus et al., 2022). Lower frequencies were indicative for more relaxed states and sleep, while higher frequencies were indicative for more aroused or stressed states (Jun & Smitha, 2016; Palatinus et al., 2022). Seet et al. (2020) applied EEG, together with subjective self-reported ratings of trust. During the experiment, artificial malfunction during the autonomous driving was initiated, during which a self-reported decrease in trust corresponded with a desire to re-engage and an increase in decision-making activity in the brain of the subject. Respondents showed decreased alpha power in the right frontal area when they had no option to take over the control, which was interpreted as enhanced motivation for controlling the vehicle. In their experiment with a real vehicle in an airport runway, Palatinus et al. (2022) showed a higher asymmetry in the frontal alpha in the human driving condition than in the automated driving condition. The authors explained the lower asymmetry of the frontal alpha activity as an indication for lower trust during automated driving. The study of Park, Shahrdar, and Nojoumian (2018) in which respondents watched videos of cars in scenarios differing in perceived risk (e.g., car performing smoothly on highway and car driving erratically and violating rules of the road) revealed an increase in the beta / alpha ratio in scenarios that were designed to be stress-inducing.

### 2.3.5. Respiratory rate

Respiratory rate is another measure for cognitive state. It is usually measured electrically using ECG or cameras (Solaz et al., 2016). Studies have found a relationship between respondent's breathing patterns and their drowsiness, stress, or cognitive load (Kiashari et al., 2020; Mehler et al., 2009; Meteier et al., 2021; Meteier et al., 2022a; Meteier et al., 2022b; Solaz et al., 2016). Heikoop, De Winter, van Arem, and Stanton (2019) compared workload of the driver in a conventional car with the workload in an automated car, but there was no significant change in respiratory rate.

### 2.3.6. Combination of assessment methods

Various researchers have applied a combination of objective assessment methods. Walker, Wang, Martens, and Verwey (2019) collected eye tracking data and GSR to measure trust in automation. Dillen et al. (2020) studied the effect of different driving styles on passengers in the vehicle, using HR, GSR and eye movements. GSR was found to be the most significant predictor of comfort and anxiety. Beggiato et al. (2020) revealed changes in the HR, pupil diameter, and interblink time in the discomfort-inducing situations. Discomfort was also rated subjectively by a direct input device, which respondents used to indicate their discomfort. In the study by Irmak (Irmak, Pool, et al., 2021), phasic galvanic skin response was associated with motion sickness measured by the MISC scale, which was associated with a decrease in comfort. Hu et al. (2016) proposed a model to measure trust by GSR and EEG signals based on empirical data from a laboratory experiment.

## 3. The road ahead

This study reports the results of a review of assessment methods that were used to measure driver's perceptions, behavior, and cognitive and affective states during (automated) driving mostly in simulated environments. This review provides insights into more objective assessment methods that can be used for AVA by passengers. A common objective measure for AVA in other domains is the actual use measured by the frequency of use of technology (Walldén, Mäkinen, & Raisamo, 2016). As AVs are not yet widely available, this measure has been rarely applied in the field of AVA. Researchers have recently started to complement the Unified Theory of Acceptance and Use of Technology (UTAUT2) with EEG and eye-tracking data to better explain the intention to use AVs (Lukovics et al., 2023). In this study, UTAUT2 questionnaires were distributed before and after the ride, while EEG and eye tracking data were collected during the ride. The authors could show that the explanatory power of the model was highest with the addition of the EEG and eye tracking data as proclaimed measure for arousal, and concluded that UTAUT2 can be applied in combination with the physiological measurements. The approach is noteworthy, and needed to advance the field of AVA. However, the UTAUT constructs were measured before and after the ride, while the EEG and eye tracking measurements were conducted during the ride. More research is needed to investigate the relationship between the self-reported and more objective physiological and behavioral measurements.

We summarize the results of the review in the conceptual model as shown by Figure 1, which provides an overview of the relationships between the factors being pivotal for the in-vehicle experience and AVA by passengers. We recommend future research to make real-time, and continuous measurements of constructs from technology acceptance models, such as the UTAUT, and physiological and behavioral measurements. In this way, we can investigate to what extent the physiological and behavioral measurements explain and predict the self-reported acceptance of AVs and their underlying mechanisms.

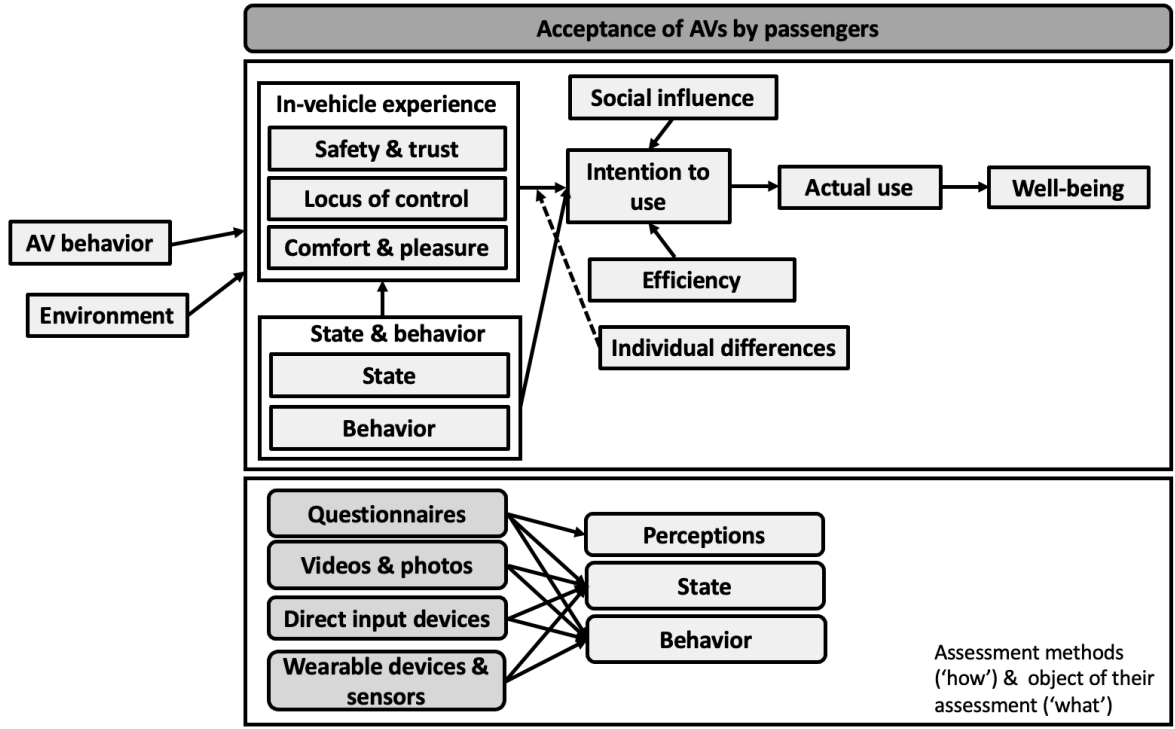

*Figure 1.* Conceptual model of acceptance of AVs by passengers (top), and assessment methods to investigate AVA by passengers (below)

## 3.1. Limitations and implications for future research

First, our work focused on instruments used to measure factors related to the driver or passenger inside AVs. The analysis of the point of view of other road users outside AVs is important for the wider public acceptance of this technology. We propose to investigate to what extent the instruments to measure passengers' in-vehicle experience and acceptance of AVs can also be applied to other road users outside AVs.

Second, most of the analyzed studies applied an experimental setup consisting of virtual reality or driving simulators. In one of our consecutive studies, we will apply a new research tool, the Autonomous Driving Passenger Experience (ADPE), to collect physiological and behavioral data from passengers in real-traffic environments in dynamic conditions, and compare them with subjective self-reported data to investigate the extent to which the physiological and behavioral indicators can serve as valid and reliable indicators for the subjective self-reported data.

**4.      Acknowledgments**

The Autonomous Driving Passenger Experience (ADPE) represents a new research tool realized by the Mobility Innovation Centre Delft (MICD), and funded by the TU Delft: Transport and Mobility Institute. The vehicle (Nissan Evalia) that constitutes the ADPE was leased to MICD by Leaseplan. The authors thank the student team that helped validating the ADPE.

This paper was submitted for presentation to the Transportation Research Board (TRB) Annual Meeting 2024.

Healey, J. A., & Picard, R. W. (2005). Detecting stress during real-world driving tasks using physiological sensors. *IEEE Transactions on Intelligent Transportation Systems, 6*, 156–166.

Heikoop, D. D., De Winter, J. C. F., van Arem, B., & Stanton, N. A. (2019). Acclimatizing to automation: Driver workload and stress during partially automated car following in real traffic. *Transportation Research Part F: Traffic Psychology and Behaviour, 65*, 503–517.

Holmes, S. R., & Griffin, M. J. (2001). Correlation between heart rate and the severity of motion sickness caused by optokinetic stimulation. *Journal of Psychophysiology, 15*, 35.

Holtgraves, T. (2004). Social desirability and self-reports: Testing models of socially desirable responding. *Personality and Social Psychology Bulletin, 30*, 161–172.

Hu, W.-L., Akash, K., Jain, N., & Reid, T. (2016). Real-time sensing of trust in human-machine interactions. *IFAC-PapersOnLine, 49*, 48–53.

Hulse, L. M., Xie, H., & Galea, E. R. (2018). Perceptions of autonomous vehicles: Relationships with road users, risk, gender and age. *Safety Science, 102*, 1–13.

Irmak, T., de Winkel, K. N., Pool, D. M., Bülthoff, H. H., & Happee, R. (2021). Individual motion perception parameters and motion sickness frequency sensitivity in fore-aft motion. *Experimental Brain Research, 239*, 1727–1745.

Irmak, T., Pool, D. M., & Happee, R. (2021). Objective and subjective responses to motion sickness: the group and the individual. *Experimental Brain Research, 239*, 515–531.

Jackson, D. N., & Messick, S. (1958). Content and style in personality assessment. *Psychological Bulletin, 55*, 243.

Jian, J.-Y., Bisantz, A. M., & Drury, C. G. (2000). Foundations for an empirically determined scale of trust in automated systems. *International Journal of Cognitive Ergonomics, 4*, 53–71.

Jun, G., & Smitha, K. G. (2016). *EEG based stress level identification.* Paper presented at the 2016 IEEE international conference on systems, man, and cybernetics (SMC).

Karjanto, J., Wils, H., Yusof, N., Terken, J., Delbressine, F., & Rauterberg, M. (2022). Measuring the perception of comfort in acceleration variation using Eletro-Cardiogram and self-rating measurement for the passengers of the automated vehicle. *J. Eng. Sci. Technol, 17*, 18–196.

Kiashari, S. E. H., Nahvi, A., Bakhoda, H., Homayounfard, A., & Tashakori, M. (2020). Evaluation of driver drowsiness using respiration analysis by thermal imaging on a driving simulator. *Multimed. Tools Appl., 79*, 17793–17815.
17